\documentclass[%
 reprint,
superscriptaddress,%groupedaddress,%unsortedaddress,%runinaddress,%frontmatterverbose,%preprint,%showpacs,preprintnumbers,%nofootinbib,%nobibnotes,%bibnotes,
amsmath,amssymb,
aps,
pra,
]{revtex4-1}

\usepackage{graphicx}% Include figure files
\usepackage{dcolumn}% Align table columns on decimal point
\usepackage{bm}% bold math

\begin{document}

\preprint{APS/123-QED}

\title{What is the maximum differential group delay \\ achievable
by a space-time wave packet in free space?}

\author{Murat Yessenov}
\author{Lam Mach}
\author{Basanta Bhaduri}
\affiliation{CREOL, The College of Optics \& Photonics, University of Central Florida, Orlando, Florida 32186, USA}

\author{Davood Mardani}
\affiliation{Dept. of Electrical and Computer Engineering, University of Central Florida, Orlando, Florida 32816, USA}

\author{H. Esat Kondakci}
\affiliation{CREOL, The College of Optics \& Photonics, University of Central Florida, Orlando, Florida 32186, USA}

\author{George K. Atia}
\affiliation{Dept. of Electrical and Computer Engineering, University of Central Florida, Orlando, Florida 32816, USA}

\author{Miguel A. Alonso}
\affiliation{The Institute of Optics, University of Rochester, Rochester, NY 14627, USA}
\affiliation{Aix Marseille Univ., CNRS, Centrale Marseille, Institut Fresnel, UMR 7249, 13397 Marseille Cedex 20, France}

\author{Ayman F. Abouraddy}
 \email{Corresponding author: raddy@creol.ucf.edu}
 \affiliation{CREOL, The College of Optics \& Photonics, University of Central Florida, Orlando, Florida 32186, USA}

%% To be edited by editor
% \dates{Compiled \today}

%\ociscodes{(140.3490) Lasers, distributed feedback; (060.2420) Fibers, polarization-maintaining;(060.3735) Fiber Bragg gratings.}

%% To be edited by editor
% \doi{\url{http://dx.doi.org/10.1364/XX.XX.XXXXXX}}

\begin{abstract}
The group velocity of `space-time' wave packets -- propagation-invariant pulsed beams endowed with tight spatio-temporal spectral correlations -- can take on arbitrary values in free space. Here we investigate theoretically and experimentally the maximum achievable group delay that realistic finite-energy space-time wave packets can achieve with respect to a reference pulse traveling at the speed of light. We find that this delay is determined \textit{solely} by the spectral uncertainty in the association between the spatial frequencies and wavelengths underlying the wave packet spatio-temporal spectrum -- and not by the beam size, bandwidth, or pulse width. We show experimentally that the propagation of space-time wave packets is delimited by a spectral-uncertainty-induced `pilot envelope' that travels at a group velocity equal to the speed of light in vacuum. Temporal walk-off between the space-time wave packet and the pilot envelope limits the maximum achievable differential group delay to the width of the pilot envelope. Within this pilot envelope the space-time wave packet can locally travel at an arbitrary group velocity and yet not violate relativistic causality because the leading or trailing edge of superluminal and subluminal space-time wave packets, respectively, are suppressed once they reach the envelope edge. Using pulses of width $\sim4$~ps and a spectral uncertainty of $\sim20$~pm, we measure maximum differential group delays of approximately $\pm150$~ps, which exceed previously reported measurements by at least three orders of magnitude. 
\end{abstract}

%\setboolean{displaycopyright}{true}

\maketitle

\section{Introduction}

Ever since Brittingham proposed in 1983 a pulsed optical beam that is transported rigidly in free space at a group velocity equal to the speed of light $c$ \cite{Brittingham83JAP}, there has been significant interest in the study of propagation-invariant wave packets \cite{Reivelt00JOSAA,Porras03PRE,Porras03PRE2,Zapata06OL,Porras17OL,Wong17ACSP1,Wong17ACSP2,Efremidis17OL,PorrasPRA18}. A variety of examples have been identified \cite{FigueroaBook8,Turunen10PO,FigueroaBook14} whose group velocity in free space -- intriguingly -- take on arbitrary values. Such pulsed optical fields are endowed with tight spatio-temporal spectral correlations \cite{Donnelly93PRSLA,Longhi04OE,Saari04PRE}, whereby each spatial frequency underlying the beam spatial structure is associated with a \textit{single} wavelength, and we hence refer to them as `space-time' (ST) wave packets \cite{Kondakci16OE,Parker16OE}. Although there is no fundamental \textit{theoretical} limit on the achievable group velocity using this strategy, previous experimental realizations -- whether subluminal or superluminal -- have not produced values that differ substantially from $c$. Indeed, the measured deviations have typically been within $\sim0.1\%$ of $c$ \cite{Bonaretti09OE,Bowlan09OL,Kuntz09PRA,Lohmus12OL,Piksarv12OE}. These experiments have recorded differential group delays on the order of 10's or 100's of femtoseconds with respect to a reference pulse traveling at $c$. This state of affairs has remained without a clear justification of the vast gap between theory and experiment.

We have recently introduced a novel spatio-temporal synthesis methodology for the preparation of ST wave packets that finally enables the full exploitation of their unique properties \cite{Kondakci17NP}. Utilizing this strategy, we have prepared ST wave packets having arbitrary group velocities in free space from $30c$ to $-4c$ \cite{Kondakci18unpub} or having a group velocity $c$ in non-dispersive optical materials independently of the refractive index \cite{Bhaduri19Optica}, in addition to synthesizing non-accelerating Airy ST wave packets \cite{Kondakci18PRL} and confirming their diffraction in time \cite{Porras17OL}, verifying self-healing \cite{Kondakci18OL}, and demonstrating extended propagation distances \cite{Bhaduri18OE,Bhaduri19unpublished} and tilted-pulse fronts \cite{Kondakci19ACSP}. An ideal ST wave packet propagates invariantly for indefinite distances, and thus can accrue in principle an arbitrary differential group delay (DGD), but requires infinite energy. Of course, only finite-energy realizations of ST wave packets are accessible experimentally, whereupon the propagation distance and the DGD become finite. We pose here the following question: what is the maximum DGD that a \textit{finite-energy} ST wave packet can achieve?

In realistic finite-energy ST wave packets, a spatio-temporal \textit{spectral uncertainty} arises in the association between the spatial frequencies and wavelengths \cite{Kondakci16OE} -- an unavoidable `fuzziness' in their association arising in any finite system \cite{Kondakci19OL}. The constraints imposed by this spectral uncertainty have not been sufficiently appreciated to date, especially in experimental realizations of ST wave packets. Traditionally, other features of a ST wave packet, such as the transverse beam size, pulse width, or bandwidth have been taken to underpin the propagation characteristics. Indeed, no previous experiment on the synthesis of propagation-invariant ST wave packets has reported the value of the spectral uncertainty.  

Here we show theoretically and experimentally that the maximum DGD of ST wave packets is determined \textit{solely} by the spectral uncertainty -- independently of beam size, pulse width, bandwidth, or ratio of the bandwidth to the spectral uncertainty. We find that the propagation \textit{distance} of a ST wave packet is determined by the spectral uncertainty and the difference between its group velocity and $c$. A theoretical model shows that finite-energy ST wave packets -- whether superluminal or subluminal -- are a product of an ideal ST wave packet (that can travel at an arbitrary group velocity) and a broad `pilot envelope' (that travels at $c$) whose width is inversely proportional to the spectral uncertainty. Temporal walk-off thus limits the distance over which arbitrary group velocities can be realized and concomitantly limits the DGD. The pilot envelope prevents the violation of relativistic causality by suppressing the leading edge of superluminal ST wave packets when approaching the envelope edge, whereas subluminal ST wave packets are suppressed at the opposite edge. Our theoretical results agree with a very recent study by Porras \cite{PorrasPRA18}.

Interferometric ultrafast pulse measurements then confirm the limits on DGD and propagation distance, and provide direct evidence for the existence of the pilot envelope by observing the predicted asymmetric suppression of superluminal and subluminal ST wave packets. We observe a DGD on the order of $\pm150$~ps for pulses of width $\sim\!4$~ps, representing a delay-bandwidth product of $\sim\!35$. This record-high observed DGD value is at least three orders-of-magnitude larger than the best previously reported results \cite{Bowlan09OL,Lohmus12OL} (4 orders-of-magnitude larger than in \cite{Giovannini15Science}), which is enabled by reducing the spectral uncertainty to $\sim\!20$~pm. Furthermore, these large DGD values are recorded over propagation distances as short as 10~mm, compared to $\sim10$~cm in \cite{Bowlan09OL,Lohmus12OL} and $\sim1$~m in \cite{Giovannini15Science}. These experimental results therefore lend support to the potential utility of ST wave packets in realizing free-space delay lines and optical buffers \cite{Zapata08JOSAA,Alfano16}.

\section{Theory}

\subsection{Ideal, infinite-energy ST wave packets}

We start from a generic wave packet $E(x,z,t)\!=\!e^{i(k_{\mathrm{o}}z-\omega_{\mathrm{o}}t)}\psi(x,z,t)$ and expand its envelope into plane waves,
\begin{equation}\label{Eq:GeneralEnvelope}
\psi(x,z,t)=\iint\!dk_{x}d\Omega\,\,\widetilde{\psi}(k_{x},\Omega)\,e^{i(k_{x}x+(k_{z}-k_{\mathrm{o}})z-\Omega t)};
\end{equation}
where the spatio-temporal spectrum $\widetilde{\psi}(k_{x},\Omega)$ is the Fourier transform of $\psi(x,0,t)$, $\omega_{\mathrm{o}}$ is the carrier frequency, $\Omega\!=\!\omega-\omega_{\mathrm{o}}$ is the frequency with respect to $\omega_{\mathrm{o}}$, $k_{\mathrm{o}}\!=\!\omega_{\mathrm{o}}/c$, and $k_{x}$ and $k_{z}$ are the transverse and longitudinal components of the wave vector along the $x$ and $z$ coordinates, respectively (we hold the field uniform along $y$). To treat space and time symmetrically, we refer to $k_{x}$ as the spatial frequency, and to $\Omega$ as the temporal frequency. An ideal ST wave packet is endowed with perfect spatio-temporal spectral correlations: each spatial frequency is associated with one temporal frequency, $\widetilde{\psi}(k_{x},\Omega)\!\rightarrow\!\widetilde{\psi}(k_{x})\delta(\Omega-\Omega(k_{x}))$, where $\Omega(k_{x})$ is a conic section resulting from the intersection of the light-cone $k_{x}^{2}+k_{z}^{2}\!=\!(\tfrac{\omega}{c})^{2}$ with a plane that is parallel to the $k_{x}$-axis and makes an angle $\theta$ (the spectral tilt angle) with the $k_{z}$-axis \cite{Yessenov18PRA} defined as $\Omega/c\!=\!(k_{z}-k_{\mathrm{o}})\tan{\theta}$. With the assumption of a delta-function correlation between spatial and temporal frequencies, the envelope in Eq.~\ref{Eq:GeneralEnvelope} takes the form
\begin{equation}\label{Eq:IdealEnvelope}
\begin{split}
\psi(x,z,t)=\int\!\!dk_{x}\,\widetilde{\psi}(k_{x})\,e^{ik_{x}x}e^{-i\Omega(t-z/v_{\mathrm{g}})}= \\ \psi(x,0,t-z/v_{\mathrm{g}}),
\end{split}
\end{equation}
where the group velocity $v_{\mathrm{g}}$ is determined by the spectral tilt angle, $v_{\mathrm{g}}\!=\!c\tan{\theta}$. Under these idealistic assumptions, the ST envelope is propagation-invariant and travels \textit{indefinitely} at a group velocity $v_{\mathrm{g}}$, such that an \textit{arbitrary} DGD can be achieved. For small bandwidths $\Delta\Omega\!\ll\!\omega_{\mathrm{o}}$, $\Omega(k_{x})$ can be approximated by a parabola \cite{Kondakci18unpub},
\begin{equation}\label{Eq:SpatialTemporalFreqRelation}
\frac{\Omega(k_{x})}{\omega_{\mathrm{o}}}=-f(\theta)\frac{k_{x}^{2}}{2k_{\mathrm{o}}^{2}}    
\end{equation}
where $f(\theta)\!=\!\tfrac{1}{\cot{\theta}-1}$. The spatial and temporal bandwidths $\Delta k_{x}$ and $\Delta\Omega$, respectively, are related through $\Delta\Omega/\omega_{\mathrm{o}}\!=\!|f(\theta)|(\Delta k_{x})^{2}/k_{\mathrm{o}}^{2}$. 

The envelope in Eq.~\ref{Eq:IdealEnvelope} is \textit{not} square-integrable and corresponds to an infinite energy. The group velocity here is the speed of the peak of the wave packet, and can take on arbitrary values by varying $\theta$. This does not violate special relativity because it cannot be used to transmit information at a speed higher than $c$ \cite{Shaarawi00JPA,SaariPRA18}. We will show below in detail how relativistic causality is upheld when considering realistic finite-energy ST wave packets.

\subsection{Previous work on realistic, finite-energy ST wave packets}

\subsubsection{Theoretical approaches}

It was immediately recognized after Brittingham's initial work \cite{Brittingham83JAP} that ideal propagation-invariant ST wave packets have infinite energy \cite{Sezginer85JAP}, and several theoretical approaches explored constructing finite-energy counterparts. The earliest approach was to superpose ideal ST wave packets \cite{Ziolkowski85JMP}; a second approach introduces a finite transverse spatial aperture \cite{Ziolkowski93JOSAA,Zamboni06JOSAA}; and a third strategy introduces a temporal `window' co-propagating with the ST wave packet (at a different group velocity) \cite{Besieris04OE,Porras17OL}.

From an experimental perspective, the delta-function correlation between spatial and temporal frequencies incorporated into Eq.~\ref{Eq:IdealEnvelope} is untenable in any finite system. Instead, an unavoidable finite `fuzziness' is introduced in the correlation between the spatial and temporal frequencies \cite{Kondakci16OE,Kondakci19OL}. Consequently, each spatial frequency $k_{x}$ is associated \textit{not} with a single frequency $\Omega\!=\!\Omega(k_{x})$, but instead with a narrow spectral range $\delta\Omega$ centered at $\Omega\!=\!\Omega(k_{x})$. We refer to $\delta\Omega$ as the spectral uncertainty ($\delta\lambda$ on the wavelength scale). This is not a statistical concept, and simply indicates that a finite spectral width is associated with each spatial frequency. The three theoretical approaches listed above all effectively relax the delta-function correlation and introduce a spectral uncertainty into the spatio-temporal spectrum of the ST wave packet. We show below that introducing a spectral uncertainty into Eq.~\ref{Eq:IdealEnvelope} leads naturally to the emergence of a time-window co-propagating with the ST wave packet (but at a group velocity of $c$) that we refer to as a `pilot envelope', a name that is inspired by an analogous concept introduced by de Broglie \cite{deBroglie60Book} and Bohm \cite{Bohm52PR}. The concept of spectral uncertainty was exploited theoretically in \cite{PorrasPRA18}, leading to similar conclusions.

\subsubsection{Proposed methodologies for synthesizing ST wave packets}

There has been a wealth of theoretical and mathematical results regarding ST wave packets over the past three and a half decades (reviewed in \cite{FigueroaBook8,Turunen10PO,FigueroaBook14}). Less effort has been devoted to developing experimental synthesis strategies. Whereas \textit{spatial} structuring of the optical field has led to a variety of optical beams (e.g., orbital angular momentum modes \cite{Allen03Book} and Airy beams \cite{Siviloglou07PRL}) and \textit{temporal} structuring of pulses has revolutionized ultrafast optics \cite{Weiner00RSI}, \textit{spatio-temporal} structuring of an optical field remains a significant challenge. Early proposals for generating ST wave packets involved utilizing time-varying apertures \cite{Shaarawi95JOSAA,Shaarawi96JOSAA} or antenna arrays \cite{Ziolkowski89PRA}. Such approaches can be viable in acoustics and ultrasonics \cite{Ziolkowski89PRL,Hernandez92JASA}, but are not practical in the optical domain, and have not -- to the best of our knowledge -- been put to test. The emergence of diffraction-free Bessel beams led to an appropriation of the techniques used in their generation for the purpose of producing ST wave packets, via annular apertures in the focal plane of a spherical lens \cite{Saari97PRL} or axicons \cite{Grunwald03PRA,Bonaretti09OE,Bowlan09OL,Alexeev02PRL}. An altogether different approach exploits the phase-matching conditions inherent in many nonlinear optical interactions to enforce the spatio-temporal spectral correlations characteristic of ST wave packets \cite{Conti03PRL,DiTrapani03PRL,Faccio06PRL,Faccio07OE}. A more recently investigated methodology relies on spatio-temporal spectral \textit{filtering} whereupon an aperture is introduced into the Fourier plane to `carve' out the desired spatio-temporal-frequency pairs  \cite{Dallaire09OE,Jedrkiewicz13OE}. Although this filtering approach was proposed for propagation-invariant wave packets propagating in disperive media (having either anomalous \cite{Dallaire09OE} or normal \cite{Jedrkiewicz13OE} dispersion), it can in principle be extended to ST wave packets that are propagation-invariant in free space.

Two comments are crucial here. First, most previous experimental efforts have been directed at generating ST wave packets with extremely broad spectra (e.g., white light from a Xe-arc lamp with 3-fs correlation time in \cite{Saari97PRL}, few-cycle pulses in \cite{Grunwald03PRA,Bock09OE,Bock17OL}, and 10's of nm of bandwidth in \cite{Faccio07OE,Bonaretti09OE}). This of course leads to many practical challenges. Although many of the mathematically obtained formulas for ST wave packets (particularly focus-wave modes and X-waves) imply the need for an ultrabroad spectrum, this is not an intrinsic feature of ST wave packets \cite{Besieris04OE} -- only the existence of the appropriate tight spatio-temporal correlations are fundamental to their unique properties. In our work, we typically make use of considerably smaller bandwidths $\Delta\lambda\!\sim\!1$~nm (but broader bandwidths are possible \cite{Kondakci18OE}). Second, a feature that has been under-appreciated to date is the importance of the spectral uncertainty to the observable features of ST wave packets. A survey of the experimental literature reveals that not a single value of spectral uncertainty $\delta\lambda$ has been reported to date. The lack of appreciation of the role of $\delta\lambda$ is compounded with the pursuit of larger bandwidths. Theoretically, the impact of $\delta\lambda$ on the propagation distance was initially discussed in \cite{Kondakci16OE} and subsequently by Porras \cite{PorrasPRA18}.

We show below that \textit{the absolute value} of the spectral uncertainty $\delta\lambda$, and \textit{not} the ratio of the full bandwidth to the spectral uncertainty $\tfrac{\Delta\lambda}{\delta\lambda}$, determines the propagation distance and DGD achievable by a ST wave packet. Previous experiments have realized large $\tfrac{\Delta\lambda}{\delta\lambda}$ ratios, but the absolute values of the spectral uncertainty has remained $\delta\lambda\widetilde{>}1$~nm. Such large values, regardless of the full spectral bandwidth, put severe limits on the DGD and the propagation distance of any ST wave packet propagating at a group velocity deviating significantly from $c$. The strategy employed in our experiments relies on a spatio-temporal spectral \textit{synthesis} procedure that we recently introduced for preparing ST wave packets \cite{Kondakci17NP}. This is an efficient phase-only technique that directly encodes a prescribed spatio-temporal spectral correlation function $\Omega\!=\!\Omega(k_{x})$ by assigning the required spatial frequency to each wavelength in the spectrum of a pulsed plane wave via a spatial light modulator (SLM) \cite{Kondakci17NP,Kondakci18PRL,Kondakci18OL,Bhaduri18OE} or a phase plate \cite{Kondakci18OE,Bhaduri19unpublished}. In contrast to previous efforts, the spectral uncertainty in our approach is typically $\delta\lambda\!\sim\!20$~pm, leading to at least three orders-of-magnitude increase in the DGD with respect to past results, in addition to the possibility of observing arbitrary values of $v_{\mathrm{g}}$ .

\subsection{Introducing spectral uncertainty into a ST wave packet}

As mentioned above, the delta-function correlation cannot be realized in practice. Instead, there is an unavoidable `fuzziness' in the association between $k_{x}$ and $\Omega$ that we refer to as the spectral uncertainty $\delta\Omega$: $\widetilde{\psi}(k_{x},\Omega)\!\rightarrow\!\widetilde{\psi}(k_{x})\widetilde{h}(\Omega-\Omega(k_{x}))$, where $\widetilde{h}$ is a narrow spectral function of width $\delta\Omega$, normalized such that $\int\!d\Omega\,|\widetilde{h}(\Omega)|^{2}\!=\!1$. This decomposition only requires that $\delta\Omega\!\ll\!\Delta\Omega$, which applies to most previous results. To obtain analytic insight into the effect of the spectral uncertainty, we make use of a Gaussian spatial spectrum $\widetilde{\psi}(k_{x})\!\propto\!\exp{\{-k_{x}^{2}/2(\Delta k_{x})^{2}\!\}}$ and spectral uncertainty function $\widetilde{h}(\Omega)\!\propto\!\exp{\{-\Omega^{2}/2(\delta\Omega)^{2}\!\}}$, whereupon the intensity profile of a finite-energy ST wave packet factorizes as follows \cite{PorrasPRA18}:
\begin{equation}
I(x,z,t)=|\psi(x,z,t)|^{2}=I_{\mathrm{ST}}(x,z,t)\cdot I_{\mathrm{p}}(x,z,t),
\end{equation}
which is a product of: (1) a narrow \textit{ideal} infinite-energy ST wave packet $I_{\mathrm{ST}}(x,z,t)$ propagating at $v_{\mathrm{g}}$ and of temporal linewidth $\tau_{\mathrm{ST}}\!\sim\!1/\Delta\Omega$ on axis,
\begin{equation}
\begin{split}
I_{\mathrm{ST}}(x,z,t)=\frac{2\sqrt{\pi}\Delta k_{x}}{\sqrt{1+[\Delta\Omega(t-z/v_{\mathrm{g}})]^{2}}} \times \\ \exp{\left\{-\frac{x^{2}(\Delta k_{x})^{2}}{1+[\Delta\Omega(t-z/v_{\mathrm{g}})]^{2}}\right\}}=I_{\mathrm{ST}}(x,0,t-z/v_{\mathrm{g}});
\end{split}
\end{equation}
and (2) a broad uncertainty-induced `pilot envelope' $I_{\mathrm{p}}(x,z,t)$ of temporal linewidth $\tau_{\mathrm{p}}\!\sim\!1/\delta\Omega$ propagating at a group velocity of $c$, 
\begin{equation}
\begin{split}
I_{\mathrm{p}}(x,z,t)=2\sqrt{\pi}\,\delta\Omega\,\exp{\left\{-(t-\tfrac{z}{c})^{2}(\delta\Omega)^{2}\!\right\}}= \\ I_{\mathrm{p}}(x,0,t-\tfrac{z}{c}).
\end{split}
\end{equation}
Note that both the ideal ST wave packet $I_{\mathrm{ST}}(x,z,t)$ \textit{and} the pilot envelope $I_{\mathrm{p}}(x,z,t)$ propagate indefinitely without change. However, their temporal walk-off stemming from the difference in their group velocities ($v_{\mathrm{g}}$ for $I_{\mathrm{ST}}$ and $c$ for $I_{\mathrm{p}}$) enforces a finite propagation distance. The pilot envelope is a plane-wave pulse, and its group velocity is simply the velocity of light in the medium ($c$ in free space).

\begin{figure*}[t!]
  \begin{center}
  \includegraphics[width=16cm]{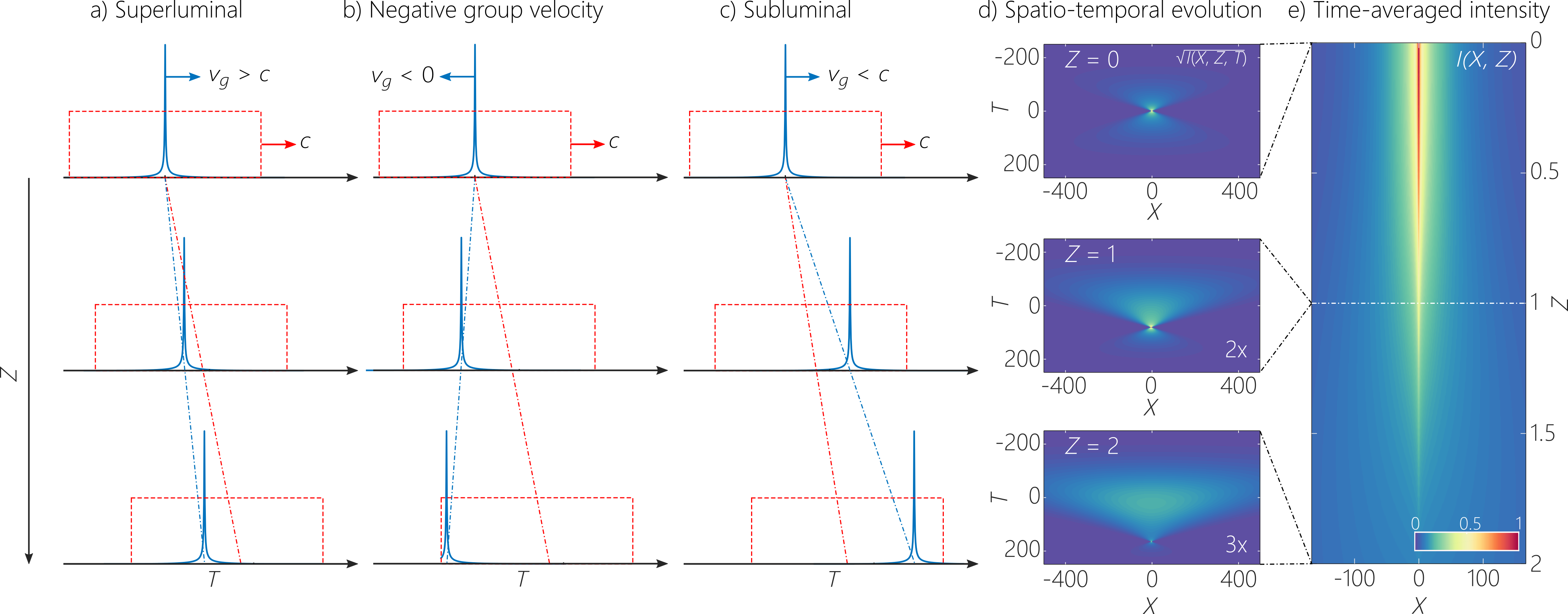}
  \end{center}
  \caption{\small{(a)-(c) Schematic illustration of finite-energy ST wave packets as the product of an ideal ST wave packet (narrow pulse, solid blue) traveling at $v_{\mathrm{g}}$ and a pilot envelope (wide pulse, dashed red) traveling at $c$: (a) superluminal $v_{\mathrm{g}}\!>\!c$; (b) negative $v_{\mathrm{g}}\!<\!0$; and (c) subluminal $v_{\mathrm{g}}\!<\!c$. In all cases, the properties of the ideal ST wave packet are maintained only over the length of the pilot envelope. (d) Snapshots of the spatio-temporal intensity profile $I(x,z,t)$ at different $z$ showing the evolution from a symmetric to an asymmetric wave packet (along the time axis) when the ST wave packet approaches the edge of the pilot envelope. The amplitudes are adjusted by the factors given in the bottom right corner for clarity. (e) The time-averaged intensity profile $I(x,z)$, showing the axial locations where the snapshots in (d) are calculated. In (d) and (e), $X$ is normalized to the inverse of the spatial bandwidth $1/\Delta k_{x}$, $T$ to $1/\Delta\Omega$, and $Z$ to the axial distance where the on-axis intensity drops to $1/e$. See also Ref.~\cite{PorrasPRA18}.}}
  \label{Fig:Concept}
\end{figure*}

This result provides conceptual clarity to several issues, as illustrated in Fig.~\ref{Fig:Concept}. It may initially appear surprising that a finite-energy ST wave packet can propagate at $v_{\mathrm{g}}\!=\!30c$ \cite{Kondakci18unpub}, for example, without violating relativistic causality. It must be recalled here that $v_{\mathrm{g}}$ refers to the velocity of the peak of the wave packet, which does not itself transmit information \cite{Smith70AJP} (see also the recent reexamination by Saari \cite{SaariPRA18}). The range over which $v_{\mathrm{g}}$ can be observed is thus the length in space (and period in time) where the ST wave packet is confined to the pilot envelope. That is, group velocities deviating from $c$ are observed only locally, delimited by the moving-window confines set by the pilot envelope that propagates at $c$. Because the ST wave packet cannot escape the confines of the pilot envelope, no information can be delivered at a speed higher than $c$, despite the reality of the propagation of the energy peak of the wave packet at an arbitrary $v_{\mathrm{g}}$. 

Because of the temporal walk-off between the ideal ST wave packet and the pilot envelope, a superluminal ST wave packet [Fig.~\ref{Fig:Concept}(a)] (or negative-$v_{\mathrm{g}}$ ST wave packet [Fig.~\ref{Fig:Concept}(b)]) is suppressed upon reaching the leading edge of the pilot envelope. A subluminal ST wave packet undergoes similar suppression when reaching the trailing edge of the pilot envelope [Fig.~\ref{Fig:Concept}(c)]. We plot in Fig.~\ref{Fig:Concept}(d) snapshots of $I(x,z,t)$ at three axial positions $z$, showing the evolution of the ST wave packet from a symmetric form when it coincides with the center of the pilot envelope, to an asymmetric form when it approaches the edge of the pilot envelope. Similar conclusions were arrived at by Porras in \cite{PorrasPRA18}.

The picture emerging here is quite distinct from that of `fast-light' traversing a resonant gain medium for example where extreme pulse-reshaping occurs accompanying strong amplification of the input pulse \cite{Boyd09Science}. For superluminal or subluminal ST wave packets, no amplification or attenuation are associated with the new group velocity. Instead, the deviation from $c$ of ST wave packets stems from the internal spatio-temporal spectral correlations introduced into the field, which also enables their propagation without distortion for potentially large distances \cite{Bhaduri18OE,Bhaduri19unpublished}. 

\subsection{Estimating the maximum differential group delay of a ST wave packet}

The maximal achievable DGD, $\tau_{\mathrm{max}}$, is thus limited by the walk-off between the ST wave packet and the pilot envelope,
\begin{equation}
\tau_{\mathrm{max}}=L_{\mathrm{max}}\left(\frac{1}{v_{\mathrm{g}}}-\frac{1}{c}\right)\sim\pm\frac{1}{\delta\Omega}\sim\pm\tau_{\mathrm{p}},
\end{equation}
where $L_{\mathrm{max}}$ is the maximum propagation distance, \begin{equation}\label{Eq:MaxDistance}
L_{\mathrm{max}}=\frac{c}{\delta\Omega}\,\frac{1}{|1-c/v_{\mathrm{g}}|}=\frac{L_{\mathrm{p}}}{|1-\cot{\theta}|};
\end{equation}
here $L_{\mathrm{p}}\!=\!c/\delta\Omega$ is the length of the pilot envelope. The positive sign is associated with subluminal wave packets, and the negative sign with superluminal wave packets. Surprisingly, $\tau_{\mathrm{max}}$ depends solely on $\delta\Omega$ and not on the beam size, pulse width, or $v_{\mathrm{g}}$, whereas $L_{\mathrm{max}}$ is determined by $|v_{\mathrm{g}}-c|$ besides $\delta\Omega$. Critically, $\tau_{\mathrm{max}}$ and $L_{\mathrm{max}}$ rely on the \textit{absolute} value of the spectral uncertainty and \textit{not} its ratio to the bandwidth as might be expected. Previous efforts featured values of the spectral uncertainty $\delta\lambda$ on the order of nanometers, thus limiting $\tau_{\mathrm{max}}$ to 10's or 100's of femtoseconds. For example, $\delta\lambda\!\sim\!1$~nm and $L_{\mathrm{max}}$ on the order of centimeters requires that $|v_{\mathrm{g}}-c|\!\sim\!10^{-4}c$, which helps explain why previous results did not realize appreciable deviations of $v_{\mathrm{g}}$ from $c$. Note that the approximation underpinning Eq.~\ref{Eq:MaxDistance} fails at $\theta\!=\!45^{\circ}$, whereupon the ST wave packet approaches a plane wave and the propagation distance grows rapidly.

\subsection{Time-averaged intensity}

The time-averaged intensity $I(x,z)\!=\!\int\!dt\,I(x,z,t)$ of a finite-energy ST wave packet can be shown to be \cite{Kondakci19OL}
\begin{equation}\label{Eq:TimeAveragedIntensity}
\begin{split}
I(x,z)\!=\!\Delta k_{x}\!\!\int_{0}^{\infty}\!\!\!ds\,\frac{e^{-s}}{\sqrt{s(1+s/\kappa^{2})}} \times\\ \exp{\left\{\!-\frac{(x\Delta k_{x}\!-\!\sqrt{s}z/z_{\mathrm{R}})^{2}}{1+s/\kappa^{2}}\!\right\}},
\end{split}
\end{equation}
where $z_{\mathrm{R}}\!=\!k_{\mathrm{o}}/(\Delta k_{x})^{2}$ is the Rayleigh range of a traditional Gaussian beam of the same spatial bandwidth as the ST wave packet, and $\kappa\!=\!\delta\Omega/\Delta\Omega$ is the ratio of the spectral uncertainty to the full bandwidth, with $\kappa\!\ll\!1$ typically. We plot $I(x,z)$ in Fig.~\ref{Fig:Concept}(e) making use of the same parameters of Fig.~\ref{Fig:Concept}(d). Note that we cannot distinguish between superluminal and subluminal wave packets from $I(x,z)$. Indeed, the role of the spectral tilt angle is only in determining the ratio of spatial to temporal bandwidths through $|f(\theta)|$, which introduces an ambiguity with respect to the sign of $f(\theta)$ that reveals whether the wave packet is subluminal or superluminal. Resolving this ambiguity requires access to the time-resolved profile $I(x,z,t)$. The on-axis intensity $I(0,z)$ from Eq.~\ref{Eq:TimeAveragedIntensity} is approximately $I(0,z)\!\approx\!\exp{\{-(\tfrac{z}{z_{\mathrm{R}}/\kappa})^{2}\}}$, so that the Rayleigh range of the ST wave packet is extended by a factor $1/\kappa$ by virtue of the spatio-temporal correlations, such that $L_{\mathrm{max}}\!\sim\!z_{\mathrm{R}}/\kappa$. Substituting for $\kappa$ and $z_{\mathrm{R}}$ we obtain the \textit{same result} in Eq.~\ref{Eq:MaxDistance}.

Therefore, two distinct physical arguments for the limit on $L_{\mathrm{max}}$ satisfyingly converge: the \textit{time-domain} argument based on walk-off between the ideal ST wave packet and the pilot envelope, and the \textit{spatial-domain} argument based on the enhancement factor in the Rayleigh range of the time-averaged intensity distribution. Furthermore, this result indicates the path forward to increasing $\tau_{\mathrm{max}}$ and $L_{\mathrm{max}}$ by realizing ever-smaller spectral uncertainty $\delta\Omega$.

\section{Interferometric measurements of the differential group delay}

\begin{figure}[t!]
  \begin{center}
  \includegraphics[width=8.6cm]{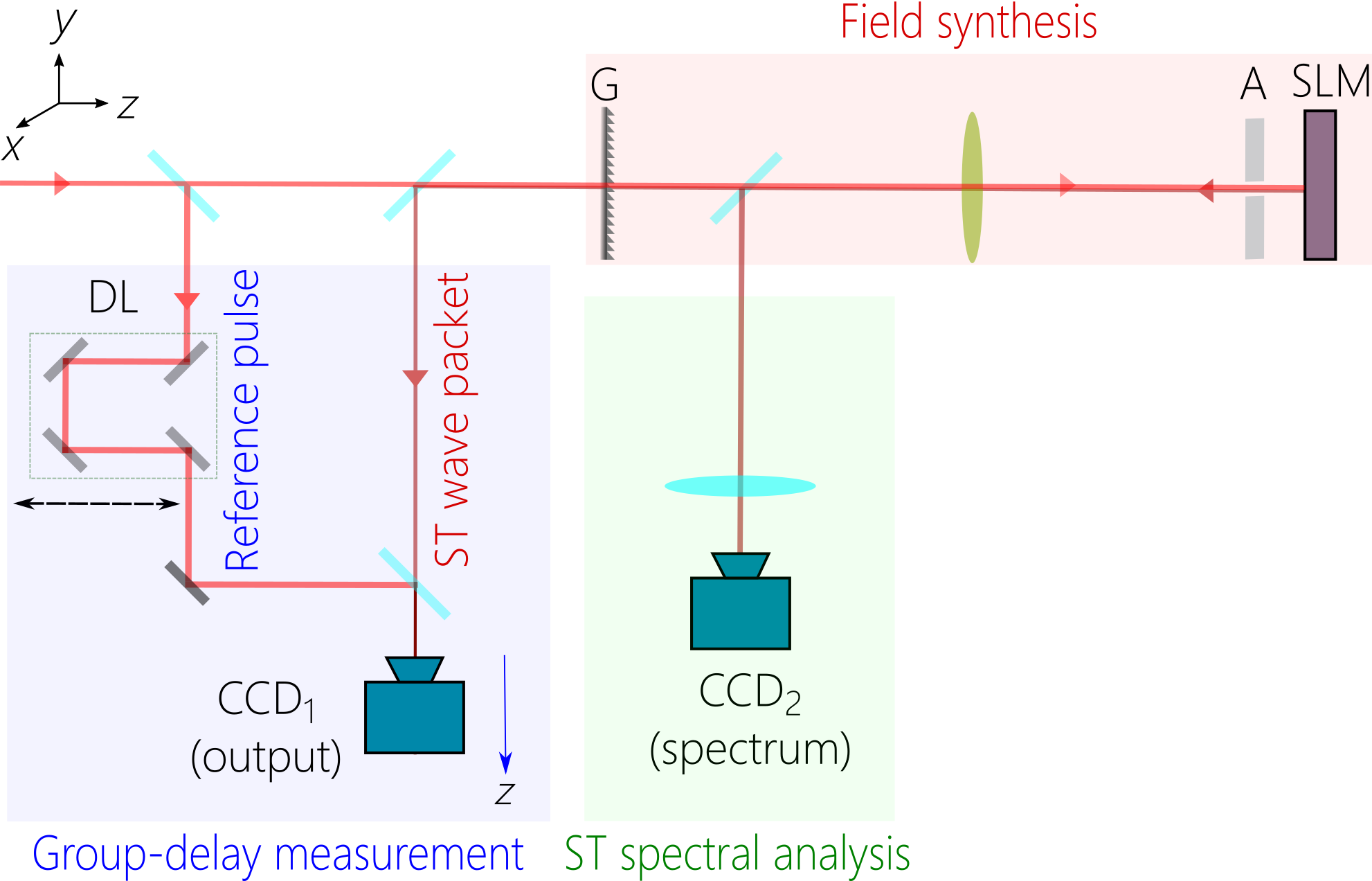}
  \end{center}
  \caption{\small{Schematic of the optical setup for synthesizing and characterizing ST wave packets. DL: Delay line; G: diffraction grating; SLM: spatial light modulator; A: aperture.}}
  \label{Fig:Setup}
\end{figure}

We now move on to the experimental realization of these theoretical predictions. Specifically, we demonstrate the impact of the pilot envelope on suppressing the leading and trailing edges of superluminal and subluminal ST wave packets, respectively, verify the dependence of $L_{\mathrm{max}}$ on $\theta$, and confirm that $\tau_{\mathrm{max}}$ is independent of $\theta$ (and thus independent of $v_{\mathrm{g}}$). A unique feature of our approach, besides its simplicity and efficiency, is its ability to synthesize ST wave packets with arbitrary group velocities that can be tuned continuously from the subluminal to superluminal regimes by only changing the phase imparted by a SLM to an incident field. Althouh the existence of luminal \cite{Brittingham83JAP}, superluminal \cite{Recami03IEEEJSTQE}, and subluminal \cite{Zamboni08PRA} ST wave packets is well-established theoretically, it was thought that different experimental configurations are needed to synthesize each \cite{Reivelt02PRE,Zapata06OL,Valtna07OC}.  

We synthesize the ST wave packets utilizing the setup established in our previous work \cite{Kondakci17NP,Kondakci18unpub,Bhaduri19Optica}, whereupon a SLM modulates the spatially resolved spectrum of a pulse in the direction orthogonal to the spectrum to assign the required $k_{x}$ to each $\lambda$. The setup is illustrated schematically in Fig.~\ref{Fig:Setup}. Starting with a generic femtosecond pulsed laser (central wavelength $\lambda_{\mathrm{o}}\!\approx\!799$~nm), we spread its spectrum spatially using a diffraction grating and collimate the spectrum with a cylindrical lens before impinging on the SLM. Each column of the SLM active area upon which wavelength $\lambda$ is incident is modulated with the appropriate spatial frequency pair $\pm k_{x}$, such that the assignment $\lambda(k_{x})$ realizes the relationship in Eq.~\ref{Eq:SpatialTemporalFreqRelation} for a prescribed $\theta$. The SLM retro-reflects the modulated wave front and the diffraction grating reconstitutes the pulse, thus forming the ST wave packet. 

Three classes of measurements are performed to charaterize each wave packet. First, we acquire the spatio-temporal spectral intensity $|\widetilde{\psi}(k_{x},\lambda)|^{2}$ after taking spatial and temporal Fourier transforms of the wave front retro-reflected from the SLM. This allows us to confirm the curvature of the spatio-temporal spectrum projected onto the $(k_{x},\lambda)$ plane, which is related to $\theta$, in addition to the spectral projection onto the $(k_{z},\tfrac{\omega}{c})$-plane. The fidelity of the modulated wave front to the prescribed ST wave packet is confirmed if the $(k_{z},\tfrac{\omega}{c})$-projection is a straight line tilted by the target $\theta$ with respect to the $k_{z}$-axis. Second, we obtain the axial evolution of the time-averaged intensity profile $I(x,z)\!=\!\int\!dt|\psi(x,z,t)|^{2}$ by scanning a CCD camera along the propagation axis. This measurement is used to obtain the propagation distance $L_{\mathrm{max}}$, which we take to be the axial distance after which the on-axis peak intensity drops by half. Third, we measure the spatio-temporal intensity profile $I(x,z,t)\!=\!|\psi(x,z,t)|^{2}$ at different axial positions $z$ through interference with a generic short reference pulse from the initial laser \cite{Kondakci18unpub,Bhaduri19Optica}. By bringing together the ST wave packet with the reference plane-wave pulse, spatially resolved interference fringes are observed when they overlap in space and time. By sweeping a delay placed in the path of the reference pulse, the decay of the visibility of the interference fringes around the ST wave packet center allows us to map out its spatio-temporal intensity profile (see \cite{Kondakci18unpub} for details). Finally, the group delay accrued by the ST wave packet as it propagates in free space can be assessed by the same interferometric technique. The maximum DGD, $\tau_{\mathrm{max}}$, is taken to be the measured group delay with respect to the reference pulse at an axial distance of $L_{\mathrm{max}}$.

The group velocity of the ST wave packet is estimated from the curvature of the spatio-temporal projection on the $(k_{x},\lambda)$-plane, or from the slope of the spectral projection onto the $(k_{z},\tfrac{\omega}{c})$-plane with respect to the $k_{z}$-axis. The group delay can then be obtained from the measured values of $L_{\mathrm{max}}$ via $\tau_{\mathrm{max}}\!=\!L_{\mathrm{max}}(\cot{\theta}-1)/c$.

\begin{figure}[t!]
  \begin{center}
  \includegraphics[width=8.6cm]{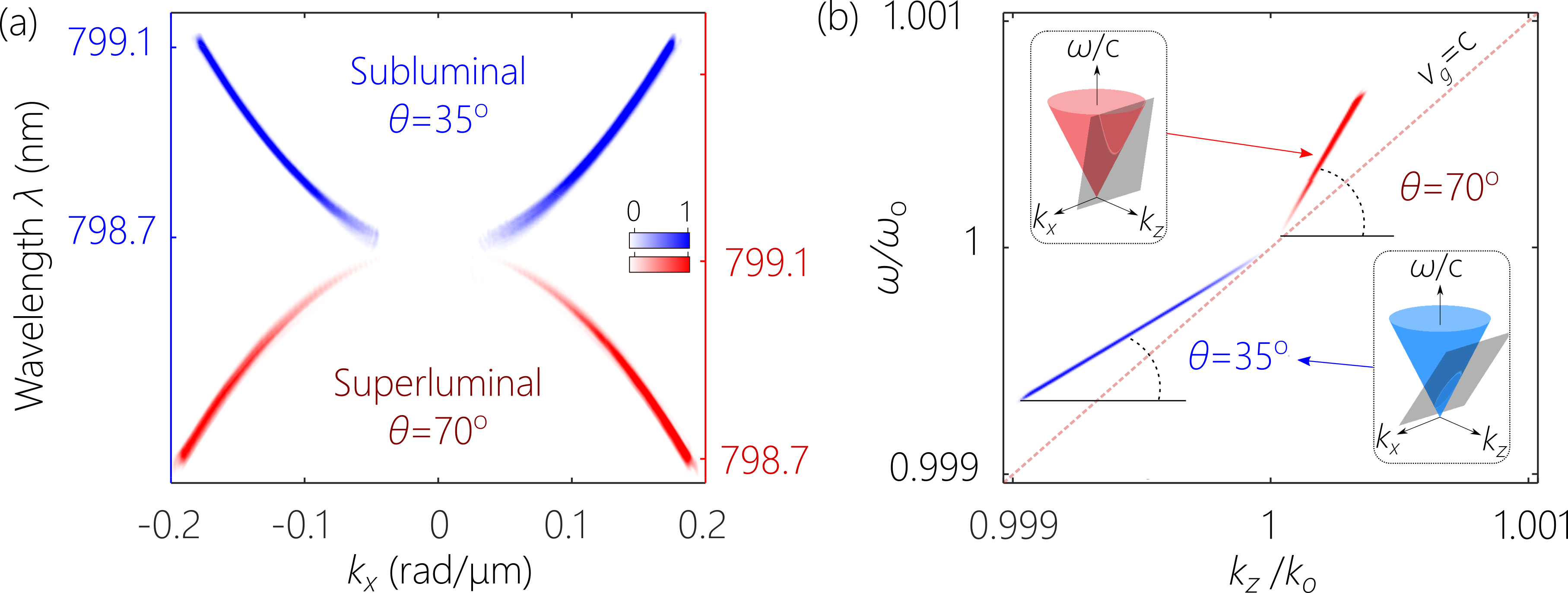}
  \end{center}
  \caption{\small{(a) Measured spatio-temporal spectral intensity $|\widetilde{\psi}(k_{x},\lambda)|^{2}$ for a subluminal ($\theta\!=\!35^{\circ}$; an ellipse) and superluminal ($\theta\!=\!70^{\circ}$; a hyperbola) ST wave packets. Both spectra appear approximately as parabolas because of the limited bandwidth ($\Delta\lambda\!\approx\!0.4$~nm), and they are shifted vertically with respect to each other for clarity. (b) Measured spatio-temporal spectra in (a) projected onto the $(k_{z},\tfrac{\omega}{c})$-plane are tilted straight lines. The insets illustrate the intersection of the tilted spectral planes planes with the light-cone.}}
  \label{Fig:Spectra}
\end{figure}

\begin{figure}[t!]
  \begin{center}
  \includegraphics[width=8.6cm]{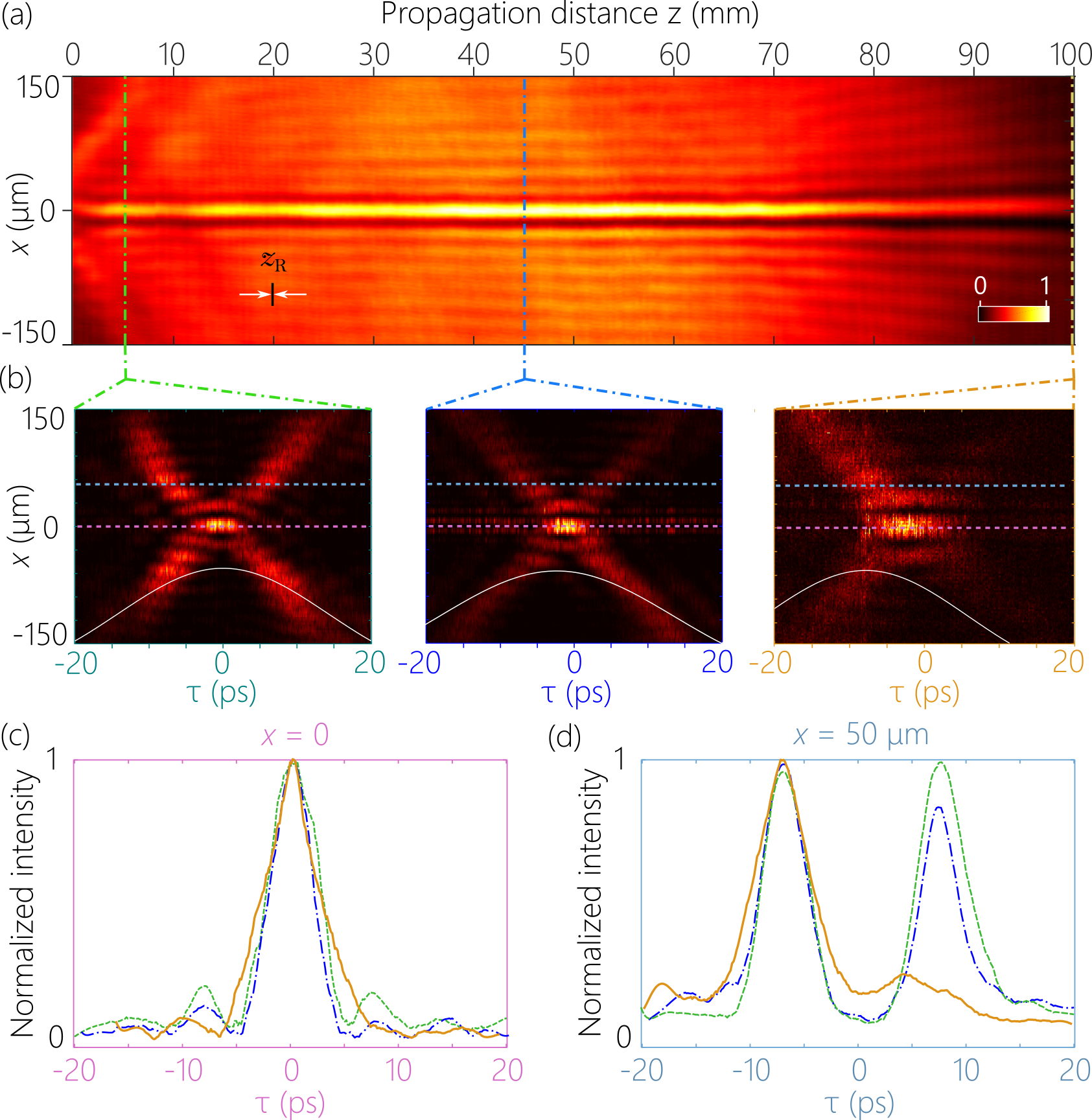}
  \end{center}
  \caption{\small{(a) Time-averaged intensity $I(x,z)$ of the subluminal ST wave packet $\theta\!=\!35^{\circ}$. (b) Time-resolved intensity $I(x,z,\tau)$ at $z\!=\!0$, 45, 100~mm. The white curves are the intensity profiles of the pilot envelope. (c) Normalized intensity at the center of the spatial profile $I(x\!=\!0,z,\tau)$ for the axial locations in (b). (d) Same as (c) at $x\!=\!50$~$\mu$m.}}
  \label{Fig:Data1}
\end{figure}

\begin{figure}[t!]
  \begin{center}
  \includegraphics[width=8.6cm]{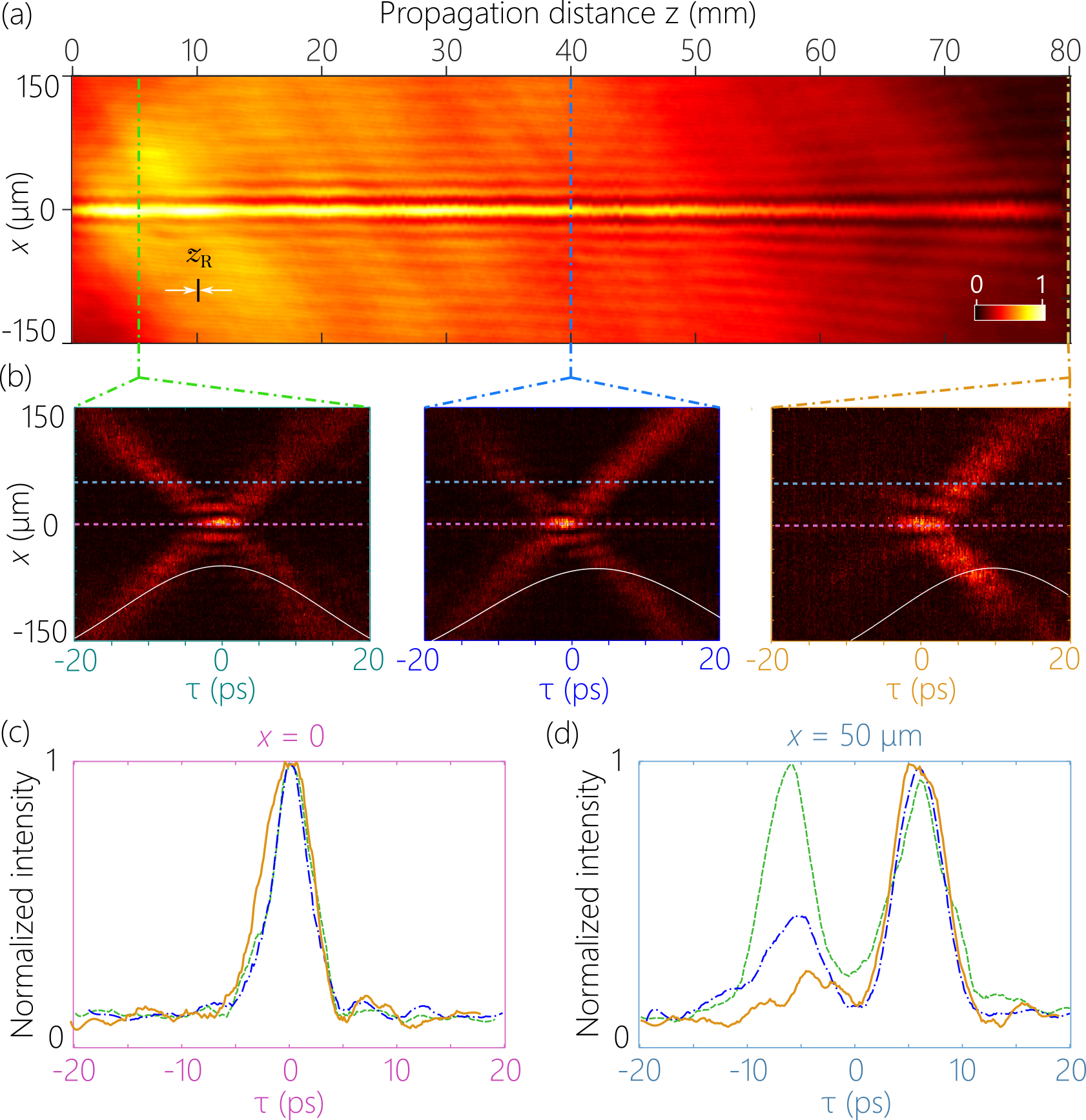}
  \end{center}
  \caption{\small{Same as Fig.~\ref{Fig:Data1} for a superluminal ST wave packet $\theta\!=\!70^{\circ}$.}}
  \label{Fig:Data2}
\end{figure}

\section{Measurement results}

\subsection{Spatio-temporal spectral measurements}

For sake of comparison, we first synthesizie two ST wave packets, a subluminal wave packet with $\theta\!=\!35^{\circ}$ ($v_{\mathrm{g}}\!\approx\!0.7c$) and a superluminal wave packet $\theta\!=\!70^{\circ}$ ($v_{\mathrm{g}}\!\approx\!2.75c$). We maintain the temporal bandwidth of each at $\Delta\lambda\!\approx\!0.4$~nm, so that the pulse width at the center of the beam ($\approx\!4.3$~ps) is the same for both. However, because of the difference in $|f(\theta)|$, their spatial bandwidths (and hence transverse spatial widths at the pulse center) are not identical.

We first plot in Fig.~\ref{Fig:Spectra}(a) the spatio-temporal spectral intensity $\widetilde{\psi}(k_{x},\lambda)$ for the subluminal ST wave packet ($\theta\!=\!35^{\circ}$) and the superluminal ST wave packet ($\theta\!=\!70^{\circ}$). Note that the sign of the curvature of the two spectra are different as determined by $f(\theta)$ (which changes sign around the luminal limit $\theta\!=\!45^{\circ}$). In the subluminal case, higher spatial frequencies are associated with larger wavelengths, whereas in the superluminal case they are associated with smaller wavelengths. This can be easily understood by examining the intersections of the spectral planes $\mathcal{P}(\theta)$ with the light-cone, as illustrated in Fig.~\ref{Fig:Spectra}(b) insets. Note that the conic section associated with the subluminal wave packet is an ellipse, whereas that associated with the superluminal wave packet is a hyperbola. However, both are well approximated by a parabola in light of the narrow bandwidth utilized.

Starting from the measured spatio-temporal spectra in the $(k_{x},\lambda)$-plane plotted in Fig.~\ref{Fig:Spectra}(a), we obtain the corresponding spatio-temporal spectra projected onto the $(k_{z},\tfrac{\omega}{c})$-plane through the free-space relationship $k_{z}^{2}\!=\!(\tfrac{\omega}{c})^{2}-k_{x}^{2}$ and plot the results in Fig.~\ref{Fig:Spectra}(b). The spectra for these two ST wave packets are straight lines tilted with respect to the $k_{z}$-axis by $35^{\circ}$ and $70^{\circ}$ as expected. Although the temporal bandwidths $\Delta\lambda$ of the two wave packets are equal and there spatial bandwidths are also close, the widths along the $k_{z}$-axis differ substantially between the subluminal and superluminal cases. 

\subsection{Measurements of the axial evolution of the time-averaged intensity}

The axial evolution of the time-averaged intensity profiles $I(x,z)$ for these two wave packets are shown in Fig.~\ref{Fig:Data1}(a) and Fig.~\ref{Fig:Data2}(a), from which we obtain $L_{\mathrm{max}}$. The slight differences in $|f(\theta)|$ for $\theta\!=\!35^{\circ}$ and $70^{\circ}$ result in a difference between the spatial widths of the central peak: the beam width is $\Delta x\!\approx\!15.6$~$\mu$m for the subluminal ST wave packet and $\Delta x\!\approx\!14$~$\mu$m for the superluminal one. This also entails a slight difference in $L_{\mathrm{max}}$ for these two cases. However, as noted earlier, we cannot determine the \textit{sign} of $f(\theta)$ from these time-averaged intensity measurements, and thus cannot distinguish the subluminal and superluminal identities of the two wave packets.

\subsection{Measurements of the time-resolved wave packet intensity profile}

The distinction between the subluminal and superluminal nature of the two ST wave packets is revealed by obtaining the time-resolved wave packet profiles $I(x,z,\tau)$, which are plotted in Fig.~\ref{Fig:Data1}(b) and Fig.~\ref{Fig:Data2}(b). Crucially, measuring $I(x,z,\tau)$ along the $z$-axis reveals the impact of the pilot envelope on either the leading or trailing edge of the ST wave packet. We obtain $I(x,z,\tau)$ at three positions for each wave packet. First, at $z\!=\!0$ the profile is symmetric and the centers of the pilot envelope and the underlying ideal ST wave packet coincide [left panels in Fig.~\ref{Fig:Data1}(b) and Fig.~\ref{Fig:Data2}(b)]. Second, at $z\!\sim\!L_{\mathrm{max}}/2$ the profile shows a slight asymmetry as temporal walk-off results in the ST wave packet approaching the pilot-envelope edge [middle panels in Fig.~\ref{Fig:Data1}(b) and Fig.~\ref{Fig:Data2}(b)]. Third, at $z\!=\!L_{\mathrm{max}}$ one edge of the wave packet is completely suppressed by the pilot envelope. This is brought out clearly be examining the temporal intensity profiles at the center of the beam $x\!=\!0$ [Fig.~\ref{Fig:Data1}(c) and Fig.~\ref{Fig:Data2}(c)] and off center $x\!=\!50$~$\mu$m [Fig.~\ref{Fig:Data1}(d) and Fig.~\ref{Fig:Data2}(d)]. It is clear that opposite sides of the superluminal and subluminal ST wave packets are suppressed (leading edge in the former, trailing edge in the latter), providing clear evidence of the existence of the pilot envelope.

\begin{figure}[t!]
  \begin{center}
  \includegraphics[width=8.6cm]{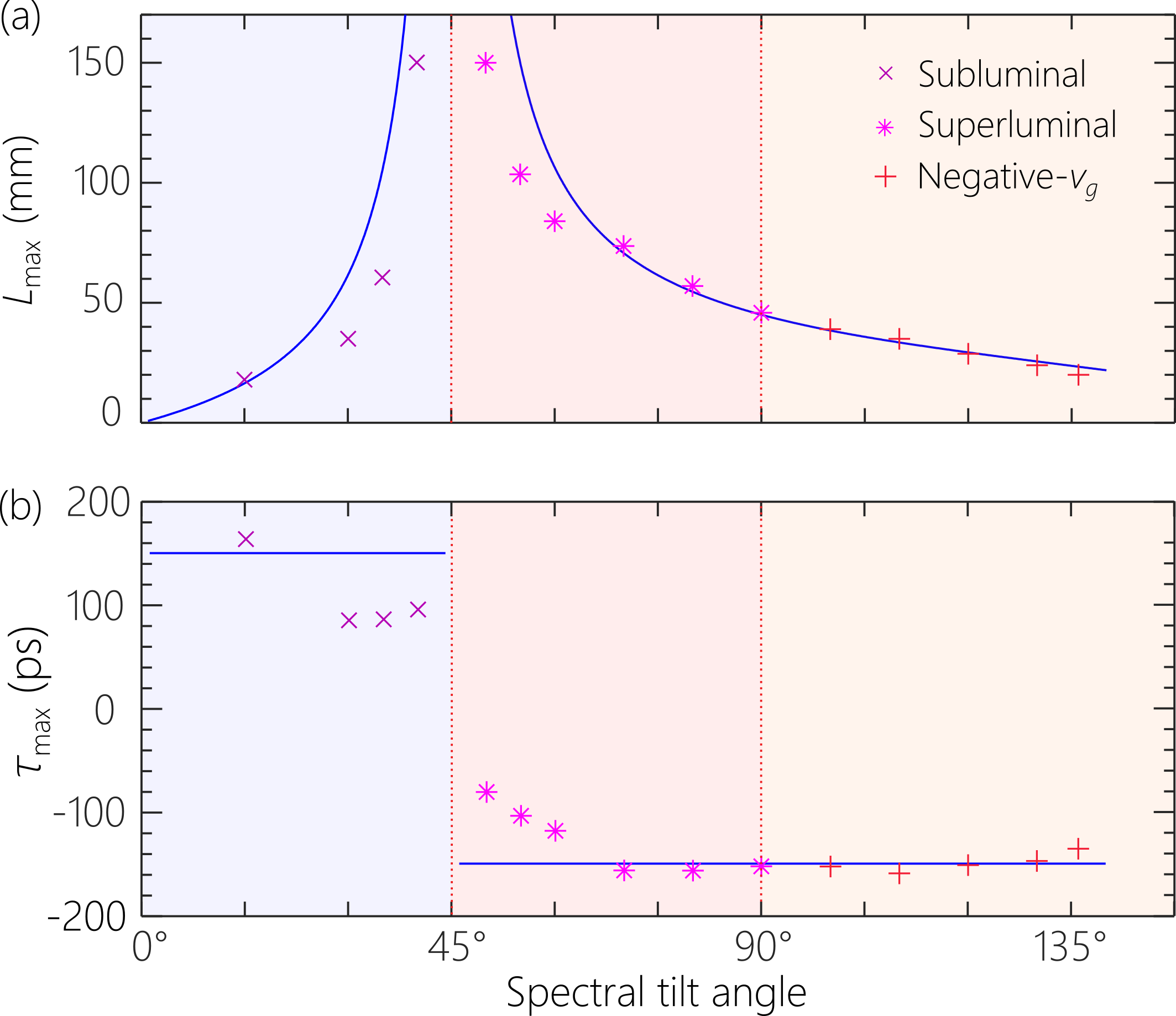}
  \end{center}
  \caption{\small{(a) Measured $L_{\mathrm{m}}$ for different $\theta$ at fixed $\delta\lambda$. Theoretical curve is $L_{\mathrm{max}}\!=\!L_{\mathrm{p}}/|1-\cot{\theta}|$, with $L_{\mathrm{p}}\!=\!45$~mm. (b) Maximum DGD as a function of $\theta$ at fixed $\delta\lambda$.}}
  \label{Fig:SpectralUncertaintyAndTheta}
\end{figure}

\subsection{Measurements of the maximum differential group delay}

The measurements of $L_{\mathrm{max}}$ while changing $\theta$ are plotted in Fig.~\ref{Fig:SpectralUncertaintyAndTheta}(a).  We vary $\theta$ in the range $15^{\circ}\!<\!\theta\!<\!135^{\circ}$, which encompasses a subluminal regime $15^{\circ}\!<\!\theta\!<\!45^{\circ}$, a superluminal regime $45^{\circ}\!<\!\theta\!<\!90^{\circ}$, and a negative-$v_{\mathrm{g}}$ regime $90^{\circ}\!<\!\theta\!<\!135^{\circ}$. The theoretical result in Eq.~\ref{Eq:MaxDistance} agrees with the data except in the vicinity of $\theta\!\rightarrow\!45^{\circ}$ where the model underlying Eq.~\ref{Eq:MaxDistance} features a singularity whereupon the ST wave packet approaches a plane wave leading to a divergence in the propagation distance and a drop to zero for the DGD. The best fit corresponds to $L_{\mathrm{p}}\!=\!45$~mm, such that $\delta\Omega/(2\pi)\!\approx\!6.6$~GHz and $\delta\lambda\!\approx\!14.2$~pm. In our experiments, $\delta\lambda$ is mainly limited by the spectral resolving power of the diffraction grating used in spreading the pulse spectrum. We estimate $\delta\lambda\!\approx\!13.5$~nm based on the second diffraction order at $\lambda_{\mathrm{o}}\!=\!800$~nm from a grating of width 25~mm having a ruling of 1200~lines/mm. In Fig.~\ref{Fig:SpectralUncertaintyAndTheta}(b) we plot  $\tau_{\mathrm{max}}\!=\!L_{\mathrm{max}}|1-\cot{\theta}|/c$. Except in the vicinity of $\theta\!\sim\!45^{\circ}$, we obtain a constant value of $\tau_{\mathrm{max}}\!\approx\!\pm150$~ps for the subluminal and superluminal wave packets. This is the largest DGD reported for any ST wave packet in free space to date and exceeds previous results by more than three orders-of-magnitude. Increasing $\delta\lambda$ serves to decrease $L_{\mathrm{max}}$ (as demonstrated in \cite{Kondakci19OL}), and thus also decrease $\tau_{\mathrm{max}}$. The delay-bandwidth product here is thus $\sim35$.

\section{Discussion}

It is important to appreciate the role of the two spectral scales relevant to ST wave packets: the full spectral bandwidth ($\Delta\Omega$ or $\Delta\lambda$) and the spectral uncertainty ($\delta\Omega$ or $\delta\lambda$). First, note that these two scales are essentially physically independent of each other. The bandwidth can be increased by increasing the size of the phase pattern displayed on the SLM or phase plate, leading to a reduction of the ST wave packet pulse width at the beam center $I(0,0,t)$. The spectral uncertainty, on the other hand, is limited in our experiment by the size of the diffraction grating (the grating spectral resolution is related to the number of grooves covered by the incident pulse). Reducing $\delta\lambda$ by increasing the grating size would \textit{not} affect $I(0,0,t)$, but would increase the wave packet propagation distance $L_{\mathrm{max}}$ (at fixed $\theta$) and the maximum DGD $\tau_{\mathrm{max}}$. The delay-bandwidth product in our measurements (the ratio of the DGD to the pulse width) is $\sim\!35$. This is substantially larger than typical values reported in slow-light studies. It remains an open question at the moment regarding the ultimate delay-bandwidth product achievable. This requires further reducing the spectral uncertainty and simultaneously increasing the bandwidth.

We highlight here some of the unique aspects of the DGD of ST wave packets. The DGD can be produced over progressively shorter distances by reducing $\theta$. Moreover, our experimental synthesis strategy allows for tuning the group velocity symmetrically from subluminal to superluminal values, thus further increasing the DGD range accessible. This is in contrast with the typical distinction between experimental approaches that produce slow-light and fast-light \cite{Boyd09Science}. It is yet to be determined what physical phenomena can benefit from the wide variability of $v_{\mathrm{g}}$ achievable with ST wave packets. We have focused here on \textit{free-space} ST wave packets, but this approach can be extended to propagation in optical materials \cite{Bhaduri19Optica,Bhaduri19unpublished}. Finally, we note that an alternative approach to spatio-temporally structured wave packets with controllable $v_{\mathrm{g}}$ has been recently proposed \cite{SaintMarie17Optica} and realized \cite{Froula18NP}, and it would be interesting to evaluate the maximum DGD it can achieve.

\section{Conclusions}

In conclusion, we have shown that the spectral uncertainty sets the limit on the maximum DGD achievable by a ST wave packet. We have derived a formula factorizing realistic ST wave packets into the product of an ideal ST wave packet and an uncertainty-induced pilot envelope. Temporal walk-off limits the DGD to the inverse of the spectral uncertainty and the maximum propagation distance to the inverse of the product of the spectral uncertainty with the deviation of the group velocity from $c$. Our measurements revealed a DGD of $\sim\!150$~ps for pulse of width $\sim\!4$~ps at the center of the spatial profile, a value that exceeds previous measurements by at least 3 orders-of-magnitude. The recorded delay-bandwidth product is $\sim35$ and can likely be increased into the range of a few hundreds by reducing the spectral uncertainty (e.g., by using a larger diffraction grating), and reducing the pulse width (using a larger temporal bandwidth \cite{Kondakci18OE}). These findings lay the foundations of a roadmap for further developments in the synthesis of ST wave packets and their applications.

\section*{Funding}
U.S. Office of Naval Research (ONR) (N00014-17-1-2458), except for MAA. National Science Foundation (NSF) (PHY-1507278); the Excellence Initiative of Aix-Marseille University -- A*MIDEX, a French ``Investissements d'Avenir'' programme for MAA.

%\bibliography{diffraction}


\end{document}